\documentclass[12pt]{article}
\usepackage[cp1251]{inputenc}

\usepackage{amsfonts,amssymb,amsthm,mathtools}
\usepackage{amsmath}
\usepackage{icomma}
\usepackage{latexsym}
\usepackage{graphicx}
\usepackage{subfigure}
\usepackage {bm}

\usepackage{geometry} 
\begin{document}
	
	\begin{center}
		\textbf{CLOSED FOLDY-WOUTHUYSEN TRANSFORMATION FOR FERMIONS MOVING IN GAUGE-INVARIANT TIME-DEPENDENT ELECTROMAGNETIC FIELDS}
	\end{center}

	\begin{center}
		
		{V.~P.~Neznamov\footnote{vpneznamov@mail.ru, vpneznamov@vniief.ru}}\\
		
		\hfil
		{\it \mbox{	Russian Federal Nuclear Center--All-Russian Research Institute of Experimental Physics},  Mira pr., 37, Sarov, 607188, Russia} \\
	\end{center}
	

\begin{abstract}
	\noindent
	\footnotesize{Previously, we obtained closed expressions for energy operators in the 
		Foldy-Wouthuysen representation in the presence of static electric fields. 
		In this case, we also established a connection between the Foldy-Wouthuysen 
		representation and the Feynman-Gell-Mann representation. In this work, we generalize these results for the case of fermions moving in time-dependent gauge-invariant electromagnetic fields..} \\
	
	
	
\end{abstract}


\section{Introduction}	

Previously \cite{bib1}, we obtained closed expressions for 
the energy operators of an electron moving in electrostatic fields in the 
chiral Foldy-Wouthuysen (CFW) representation \cite{bib2}. 
The transition to the CFW representation was implemented using Dirac 
matrices in the chiral representation.

Closed expressions for Hamiltonians or energy operators are necessary in the 
presence of strong electromagnetic fields. In this case, perturbation theory 
is not practical, and we are in the realm of non-perturbative quantum 
electrodynamics.

In this work, we generalize the results of Ref. \cite{bib1} for 
the case of electrons moving in time-dependent gauge-invariant 
electromagnetic fields.

In Sec. 2 of the article, the Dirac equation in a dynamic gauge-invariant 
electromagnetic field is considered using Dirac matrices in the standard and 
chiral representations. In Sec. 3, closed Foldy-Wouthuysen 
transformations are performed with Dirac matrices in the chiral 
representation. In Sec. 4, the fermion equations in the Foldy-Wouthuysen 
representation in the presence of dynamic electromagnetic fields are 
obtained. A direct connection between the Foldy-Wouthuysen and 
Feynman-Gell-Mann representations is demonstrated. The Conclusion summarizes 
the main results of our work.


\section{Dirac equation for an electron in a Time-Dependent Gauge-Invariant 	Electromagnetic Field}
The Dirac equation in a time-dependent electromagnetic field has the form
\begin{equation}
	\label{eq1}
	p^{0}\psi_{D} ({\rm {\bf x}},t)=H_{D} (t)\,\psi_{D} ({\rm {\bf 
			x}},t)=\left( {{\rm {\bm {\alpha \pi} }}({\rm {\bf x}},t)+\beta m+eA^{0}({\rm 
			{\bf x}},t)} \right)\psi_{D} ({\rm {\bf x}},t).
\end{equation}
Hereinafter, we use the system of units of $\hslash =c=1$; $H_{D} (t)$ is the 
Dirac Hamiltonian; $\psi_{D} ({\rm {\bf x}},t)$ is the bispinor wave 
function; $p^{\mu }=i\dfrac{\partial }{\partial x_{\mu } };\,\,\,\mu 
=0,1,2,3$; ${\rm {\bm {\pi} }}({\rm {\bf x}},t)={\rm {\bf p}}-e{\rm {\bf 
		A}}({\rm {\bf x}},t)$; $A^{\mu }({\rm {\bf x}},t)$ is the electromagnetic 
four-potential; $\alpha^{i},\,\,\beta $ are four-dimensional Dirac 
matrices $\left( {i=1,2,3} \right)$.

The Dirac matrices in the standard representation have the form
\begin{equation}
	\label{eq2}
	\alpha^{i}=\left( {\begin{array}{l}
			0\;\;\,\sigma^{i}{\kern 1pt} \\ 
			\sigma^{i}\;\;0 \\ 
	\end{array}} \right),\;\,\,\beta =\gamma^{0}=\left( {\begin{array}{l}
			I\;\;\,\,\,\,\,{\kern 1pt}0 \\ 
			0\;\,-I \\ 
	\end{array}} \right),\;\,\,\gamma_{5} =\left( {\begin{array}{l}
			0\;\;\,{\kern 1pt}I \\ 
			I\;\;\;0 \\ 
	\end{array}} \right),\;\,\,\gamma^{i}=\gamma^{0}\alpha^{i}.
\end{equation}
Here, $\sigma^{i}$ are the two-dimensional Pauli matrices, $I$ is the 
two-dimensional identity matrix.

In modern gauge field theories, and in particular in the Standard Model, the 
chiral representation of Dirac matrices is widely used
\begin{equation}
	\label{eq3}
	\begin{array}{l}
		\alpha_{c}^{i} =S\alpha^{i} S^{-1}=\left( {\begin{array}{l}
				\sigma^{i}\;\;\,\,\,\,\,\,0{\kern 1pt} \\ 
				0\;\;-\;\sigma^{i} \\ 
		\end{array}} \right),\;\,\beta_{c} =\gamma_{c}^{0} =S\beta S^{-1}=\left( 
		{\begin{array}{l}
				0\;\;\,{\kern 1pt}I \\ 
				I\;\,\,0 \\ 
		\end{array}} \right), \\ 
		\,\,\,\,\,\,\,\,\,\,\,\,\,\,\,\,\,\,\,\,\,\gamma_{c}^{5} =S\gamma^{5} 
		S^{-1}=\left( {\begin{array}{l}
				I\;\;\,\,\,\,{\kern 1pt}0 \\ 
				0\;-I \\ 
		\end{array}} \right),\;\gamma_{c}^{i} =\gamma_{c}^{0} \alpha_{c}^{i} . 
		\\ 
	\end{array}
\end{equation}
In Eq. (\ref{eq3}), the unitary transformation matrix $S$ has the form
\begin{equation}
	\label{eq4}
	S=\frac{1}{\sqrt 2 }\left( {{\begin{array}{*{20}c}
				I \hfill & \,\,\,\,I \hfill \\
				I \hfill & {-I} \hfill \\
	\end{array} }} \right).
\end{equation}
The Dirac equation (\ref{eq1}) with matrices in the chiral representation (\ref{eq3}) has the form
\begin{equation}
	\label{eq5}
	p^{0}\psi_{c} ({\rm {\bf x}},t)=H_{c} (t)\,\psi_{c} ({\rm {\bf 
			x}},t)=\left( {{\rm {\bm {\alpha} }}_{c} {\rm {\bm \pi }}({\rm {\bf 
				x}},t)+\beta_{c} m+eA^{0}({\rm {\bf x}},t)} \right)\psi_{c} ({\rm {\bf 
			x}},t).
\end{equation}
Here,
\begin{equation}
	\label{eq6}
	\psi_{c} ({\rm {\bf x}},t)=S\,\psi_{D} ({\rm {\bf x}},t),
\end{equation}
\begin{equation}
	\label{eq7}
	H_{c} (t)\,=SH_{D} S^{-1}.
\end{equation}
Representation (\ref{eq3}) in Eq. (\ref{eq5}) allows us to obtain closed expressions 
for Foldy-Wouthuysen transformations in the presence of time-dependent 
gauge-invariant electromagnetic fields.

The bispinor (\ref{eq6}) can be presented as a column of two spinors
\begin{equation}
	\label{eq8}
	\psi_{c} ({\rm {\bf x}},t)=\left( {{\begin{array}{*{20}c}
				{\varphi_{c} ({\rm {\bf x}},t)} \hfill \\
				{\chi_{c} ({\rm {\bf x}},t)} \hfill \\
	\end{array} }} \right).
\end{equation}
From Eq. (\ref{eq5}) with the representation (\ref{eq3}), the following equalities can be obtained
\begin{equation}
	\label{eq9}
	p^{0}\varphi_{c} ({\rm {\bf x}},t)={\rm {\bm {\sigma \pi} }}({\rm {\bf 
			x}},t)\varphi_{c} ({\rm {\bf x}},t)+m\chi_{c} \left( {{\rm {\bf x}},t} 
	\right)+eA^{0}({\rm {\bf x}},t)\varphi_{c} ({\rm {\bf x}},t),
\end{equation}
\begin{equation}
	\label{eq10}
	p^{0}\chi_{c} ({\rm {\bf x}},t)=-{\rm {\bm {\sigma \pi} }}({\rm {\bf 
			x}},t)\chi_{c} ({\rm {\bf x}},t)+m\varphi_{c} \left( {{\rm {\bf x}},t} 
	\right)+eA^{0}({\rm {\bf x}},t)\chi_{c} ({\rm {\bf x}},t),
\end{equation}
\begin{equation}
	\label{eq11}
	\chi_{c} \left( {{\rm {\bf x}},t} \right)=\frac{m}{p^{0}+{\rm {\bm {\sigma 
				\pi }}}\left( {{\rm {\bf x}},t} \right)-eA^{0}\left( {{\rm {\bf x}},t} 
		\right)}\varphi_{c} \left( {{\rm {\bf x}},t} \right),
\end{equation}
\begin{equation}
	\label{eq12}
	\varphi_{c} \left( {{\rm {\bf x}},t} \right)=\frac{m}{p^{0}-{\rm {\bm 
				{\sigma \pi} }}\left( {{\rm {\bf x}},t} \right)-eA^{0}\left( {{\rm {\bf x}},t} 
		\right)}\chi_{c} \left( {{\rm {\bf x}},t} \right).
\end{equation}
Taking into account (\ref{eq11}) and (\ref{eq12}), the bispinor (\ref{eq8} ) can be written using only the 
spinor $\varphi_{c} \left( {{\rm {\bf x}},t} \right)$ or only the spinor $\chi 
_{c} \left( {{\rm {\bf x}},t} \right)$
\begin{equation}
	\label{eq13}
	\psi_{c}^{\left( + \right)} ({\rm {\bf x}},t)=A^{\left( + \right)}\left( 
	{{\begin{array}{*{20}c}
				\,\,\,\,\,\,\,\,\,\,\,\,\,\,\,\,\,\,\,\,{\varphi_{c} ({\rm {\bf x}},t)} \hfill \\ [10pt]
				{\dfrac{m}{p^{0}+{\rm {\bm {\sigma \pi} }}\left( {{\rm {\bf x}},t} 
						\right)-eA^{0}\left( {{\rm {\bf x}},t} \right)}\varphi_{c} ({\rm {\bf 
							x}},t)} \hfill \\
	\end{array} }} \right),
\end{equation}
\begin{equation}
	\label{eq14}
	\psi_{c}^{\left( - \right)} ({\rm {\bf x}},t)=A^{\left( - \right)}\left( 
	{{\begin{array}{*{20}c}
				{-\dfrac{m}{-p^{0}+{\rm {\bm {\sigma \pi} }}\left( {{\rm {\bf x}},t} 
						\right)+eA^{0}\left( {{\rm {\bf x}},t} \right)}\chi_{c} ({\rm {\bf x}},t)} 
				\hfill \\ [10pt]
				\,\,\,\,\,\,\,\,\,\,\,\,\,\,\,\,\,\,\,\,\,\,\,\,\,\,{\chi_{c} ({\rm {\bf x}},t)} \hfill \\
	\end{array} }} \right).
\end{equation}
The normalized operators $A^{\left( + \right)}$ and $A^{\left( - \right)}$ are 
found from the conditions
\begin{equation}
	\label{eq15}
	\psi_{c}^{\left( + \right)} ({\rm {\bf x}},t)^{\dag }\psi_{c}^{\left( + 
		\right)} ({\rm {\bf x}},t)=\varphi_{c}^{\dag } ({\rm {\bf x}},t)\varphi 
	_{c} ({\rm {\bf x}},t),
\end{equation}
\begin{equation}
	\label{eq16}
	\psi_{c}^{\left( - \right)} ({\rm {\bf x}},t)^{\dag }\psi_{c}^{\left( - 
		\right)} ({\rm {\bf x}},t)=\chi_{c}^{\dag } ({\rm {\bf x}},t)\chi_{c} 
	({\rm {\bf x}},t).
\end{equation}
Therefore,
\begin{equation}
	\label{eq17}
	A^{\left( + \right)} =\left( {1+\frac{m^{2}}{\left( {p^{0}+{\rm {\bm 
						{\sigma \pi} }}\left( {{\rm {\bf x}},t} \right)-eA^{0}\left( {{\rm {\bf x}},t} 
				\right)} \right)^{2}}} \right)^{-1 \mathord{\left/ {\vphantom {1 2}} \right. 
			\kern-\nulldelimiterspace} 2}\,,
\end{equation}
\begin{equation}
	\label{eq18}
	A^{\left( - \right)} =\left( {1+\frac{m^{2}}{\left( {-p^{0}+{\rm {\bm 
						{\sigma \pi} }}\left( {{\rm {\bf x}},t} \right)+eA^{0}\left( {{\rm {\bf x}},t} 
				\right)} \right)^{2}}} \right)^{-1 \mathord{\left/ {\vphantom {1 2}} \right. 
			\kern-\nulldelimiterspace} 2}\,.
\end{equation}
In the equalities (\ref{eq13}), (\ref{eq14}) and (\ref{eq17}), (\ref{eq18}), the numerators of the expressions 
commute with the corresponding denominators. This allows us to obtain closed 
expressions of Foldy-Wouthuysen transformations in the presence of dynamic 
(time-dependent) electromagnetic fields.


\section{Closed Foldy-Wouthuysen transformations with Dirac matrices in the chiral representation }

Let us recall two conditions of the transition to the Foldy-Wouthuysen 
representation (see, for instance, Ref. \cite{bib3}):

\begin{enumerate}
	\item The Hamiltonian and fermion equations in the FW representation are diagonal relative to the upper and lower spinors of the transformed wave function $\psi_{FW} ({\rm {\bf x}},t)$.
	\item In the Foldy-Wouthuysen transformations (including CFW transformations in presence of dynamic electromagnetic fields), the reduction conditions for the wave function must be satisfied.
\end{enumerate}
In our case \footnote{Below, 
	the wave functions are normalized by unitary probability in the box of volume $V$. 
	For brevity, in our expressions for wave functions, there are no multipliers 
	$1 \mathord{\left/ {\vphantom {1 {\sqrt V }}} \right. 
		\kern-\nulldelimiterspace} {\sqrt V }$.},
\begin{equation}
	\label{eq19}
	\psi_{CFW}^{\left( + \right)} ({\rm {\bf x}},t)=U_{CFW}^{\left( + \right)} 
	({\rm {\bf x}},t)\psi_{c}^{\left( + \right)} ({\rm {\bf x}},t)=\left( 
	{{\begin{array}{*{20}c}
				{\varphi_{c} ({\rm {\bf x}},t)} \hfill \\
				\,\,\,\,\,\,\,\,0 \hfill \\
	\end{array} }} \right),
\end{equation}
\begin{equation}
	\label{eq20}
	\psi_{CFW}^{\left( - \right)} ({\rm {\bf x}},t)=U_{CFW}^{\left( - \right)} 
	({\rm {\bf x}},t)\psi_{c}^{\left( - \right)} ({\rm {\bf x}},t)=\left( 
	{{\begin{array}{*{20}c}
			\,\,\,\,\,\,\,\,0 \hfill \\
				{\chi_{c} ({\rm {\bf x}},t)} \hfill \\
	\end{array} }} \right).
\end{equation}
The equalities (\ref{eq19}), (\ref{eq20}) and (\ref{eq13}), (\ref{eq14}) allow us to define the unitary operators $U_{CFW}^{\left( + \right)\,\,\,\,\,\dag }=\left( {U_{CFW}^{\left( + 
		\right)} } \right)^{-1},\,\,\,U_{CFW}^{\left( - \right)\,\,\,\,\dag }=\left( 
{U_{CFW}^{\left( - \right)} } \right)^{-1}$:
\begin{equation}
	\label{eq21}
	U_{CFW}^{\left( + \right)} =A^{\left( + \right)}\left( 
	{1+\frac{m}{p^{0}+{\rm {\bm {\sigma \pi} }}\left( {{\rm {\bf x}},t} 
			\right)-eA^{0}\left( {{\rm {\bf x}},t} \right)}\gamma_{c}^{5} \beta_{c} } 
	\right),
\end{equation}
\begin{equation}
	\label{eq22}
	U_{CFW}^{\left( - \right)} =A^{\left( - \right)}\left( 
	{1+\frac{m}{-p^{0}+{\rm {\bm {\sigma \pi} }}\left( {{\rm {\bf x}},t} 
			\right)+eA^{0}\left( {{\rm {\bf x}},t} \right)}\gamma_{c}^{5} \beta_{c} } 
	\right).
\end{equation}

\section{Fermion equations in the Foldy-Wouthuysen representation in the presence of dynamic electromagnetic fields}

The time-dependent Hamiltonians in the FW representation have the form
\begin{equation}
	\label{eq23}
	H_{CFW}^{\left( + \right)} \left( t \right)=U_{CFW}^{\left( + \right)} 
	\left( t \right)H_{c} \left( t \right)U_{CFW}^{\left( + \right)} \left( t 
	\right)^{\dag }-iU_{CFW}^{\left( + \right)} \left( t \right)\frac{\partial 
	}{\partial t}\left( {U_{CFW}^{\left( + \right)} \left( t \right)^{\dag }} 
	\right)\,,
\end{equation}
\begin{equation}
	\label{eq24}
	H_{CFW}^{\left( - \right)} \left( t \right)=U_{CFW}^{\left( - \right)} 
	\left( t \right)H_{c} \left( t \right)U_{CFW}^{\left( - \right)} \left( t 
	\right)^{\dag }-iU_{CFW}^{\left( - \right)} \left( t \right)\frac{\partial 
	}{\partial t}\left( {U_{CFW}^{\left( - \right)} \left( t \right)^{\dag }} 
	\right)\,.
\end{equation}
The second terms in (\ref{eq23}) and (\ref{eq24}) are absent in case of static electromagnetic 
fields. $H_{CFW}^{\left( + \right)} \left( t \right)$ acts on the function 
(\ref{eq19}) with
\begin{equation}
	\label{eq25}
	\gamma_{c}^{5} \psi_{CFW}^{\left( + \right)} \left( t \right)=\psi 
	_{CFW}^{\left( + \right)} \left( t \right).
\end{equation}
$H_{CFW}^{\left( - \right)} \left( t \right)$ acts on function (\ref{eq20}) with
\begin{equation}
	\label{eq26}
	\gamma_{c}^{5} \psi_{CFW}^{\left( - \right)} \left( t \right)=-\psi 
	_{CFW}^{\left( - \right)} \left( t \right).
\end{equation}
The Hamiltonians (\ref{eq23}) and (\ref{eq24}) can be written as the sum of even and odd parts
\begin{equation}
	\label{eq27}
	H_{CFW}^{\left( + \right)} \left( t \right)=\left( {H_{CFW}^{\left( + 
			\right)} \left( t \right)} \right)_{even} +\left( {H_{CFW}^{\left( + 
			\right)} \left( t \right)} \right)_{odd} ,
\end{equation}
\begin{equation}
	\label{eq28}
	H_{CFW}^{\left( - \right)} \left( t \right)=\left( {H_{CFW}^{\left( - 
			\right)} \left( t \right)} \right)_{even} +\left( {H_{CFW}^{\left( - 
			\right)} \left( t \right)} \right)_{odd} .
\end{equation}
Here, $\left( {H_{CFW}^{\left( + \right)} \left( t \right)} \right)_{odd} 
,\,\,\,\left( {H_{CFW}^{\left( - \right)} \left( t \right)} \right)_{odd} 
$ are the parts of the Hamiltonians with the matrix $\beta_{c} $, which mixes 
the upper and lower spinors of the wave function.

In the Foldy-Wouthuysen representation, according to condition 1 in Sec. 3,
\begin{equation}
	\label{eq29}
	\left( {H_{CFW}^{\left( + \right)} \left( t \right)} \right)_{odd} \psi 
	_{CFW}^{\left( + \right)} \left( {{\rm {\bf x}},t} \right)=0,
\end{equation}
\begin{equation}
	\label{eq30}
	\left( {H_{CFW}^{\left( - \right)} \left( t \right)} \right)_{odd} \psi 
	_{CFW}^{\left( - \right)} \left( {{\rm {\bf x}},t} \right)=0.
\end{equation}
When (\ref{eq29}) and (\ref{eq30}) are satisfied, the Hamiltonians are
\begin{equation}
	\label{eq31}
	H_{CFW}^{\left( + \right)} \left( t \right)=\left( {H_{CFW}^{\left( + 
			\right)} \left( t \right)} \right)_{even} ,
\end{equation}
\begin{equation}
	\label{eq32}
	H_{CFW}^{\left( - \right)} \left( t \right)=\left( {H_{CFW}^{\left( - 
			\right)} \left( t \right)} \right)_{even} .
\end{equation}
The equalities (\ref{eq31}) and (\ref{eq32}) do not contain terms with the matrix $\beta_{c} $.

From Eqs.(\ref{eq5}), (\ref{eq21}) - (\ref{eq24}), (\ref{eq27}), (\ref{eq28}), we define the 
operators $\left( {H_{CFW}^{\left( + \right)} \left( t \right)} \right)_{odd} 
$ and $\left( {H_{CFW}^{\left( - \right)} \left( t \right)} \right)_{odd} $
\begin{equation}
	\label{eq33}
	\begin{array}{l}
		\left( {H_{CFW}^{\left( + \right)} \left( t \right)} \right)_{odd} =\beta 
		_{c} mA^{\left( + \right)}\left[ {1-\dfrac{m^{2}}{\left( {p^{0}+{\rm {\bm 
							{\sigma \pi} }}\left( {{\rm {\bf x}},t} \right)-eA^{0}\left( {{\rm {\bf x}},t} 
					\right)} \right)^{2}}-} \right. \\ [15pt]
		-\dfrac{1}{p^{0}+{\rm {\bm {\sigma \pi} }}\left( {{\rm {\bf x}},t} 
			\right)-eA^{0}\left( {{\rm {\bf x}},t} \right)}\left( {-p^{0}+{\rm {\bm 
					{\sigma \pi} }}\left( {{\rm {\bf x}},t} \right)+\gamma_{c}^{5} eA^{0}\left( 
			{{\rm {\bf x}},t} \right)} \right)- \\  [15pt]
		\left. {-\left( {p^{0}+{\rm {\bm {\sigma \pi} }}\left( {{\rm {\bf x}},t} 
				\right)-\gamma_{c}^{5} eA^{0}\left( {{\rm {\bf x}},t} \right)} 
			\right)\dfrac{1}{p^{0}+{\rm {\bm {\sigma \pi} }}\left( {{\rm {\bf x}},t} 
				\right)-eA^{0}\left( {{\rm {\bf x}},t} \right)}} \right]A^{\left( + 
			\right)}, \\ 
	\end{array}
\end{equation}
\begin{equation}
	\label{eq34}
	\begin{array}{l}
		\left( {H_{CFW}^{\left( - \right)} \left( t \right)} \right)_{odd} =\beta 
		_{c} mA^{\left( - \right)}\left[ {1-\dfrac{m^{2}}{\left( {-p^{0}+{\rm {\bm 
							{\sigma \pi} }}\left( {{\rm {\bf x}},t} \right)+eA^{0}\left( {{\rm {\bf x}},t} 
					\right)} \right)^{2}}-} \right. \\ [15pt]
		-\dfrac{1}{-p^{0}+{\rm {\bm {\sigma \pi} }}\left( {{\rm {\bf x}},t} 
			\right)+eA^{0}\left( {{\rm {\bf x}},t} \right)}\left( {p^{0}+{\rm {\bm 
					{\sigma \pi} }}\left( {{\rm {\bf x}},t} \right)+\gamma_{c}^{5} eA^{0}\left( 
			{{\rm {\bf x}},t} \right)} \right)- \\ [15pt]
		\left. {-\left( {-p^{0}+{\rm {\bm {\sigma \pi} }}\left( {{\rm {\bf x}},t} 
				\right)-\gamma_{c}^{5} eA^{0}\left( {{\rm {\bf x}},t} \right)} 
			\right)\dfrac{1}{-p^{0}+{\rm {\bm {\sigma \pi} }}\left( {{\rm {\bf x}},t} 
				\right)+eA^{0}\left( {{\rm {\bf x}},t} \right)}} \right]A^{\left( - 
			\right)}. \\ 
	\end{array}
\end{equation}
Equations (\ref{eq29}), (\ref{eq30}) with the expressions (\ref{eq33}), (\ref{eq34}) and with the use of equalities (\ref{eq25}), (\ref{eq26}) are reduced to the form
\begin{equation}
	\label{eq35}
	\begin{array}{l}
		\left( {H_{CFW}^{\left( + \right)} \left( t \right)} \right)_{odd} \left( 
		{{\begin{array}{*{20}c}
					{\varphi_{c} ({\rm {\bf x}},t)} \hfill \\
					\,\,\,\,\,\,\,0 \hfill \\
		\end{array} }} \right)=\beta_{c} m\dfrac{A^{\left( + \right)}}{\left( 
			{p^{0}+{\rm {\bm {\sigma \pi} }}\left( {{\rm {\bf x}},t} \right)-eA^{0}\left( 
				{{\rm {\bf x}},t} \right)} \right)^{2}}\times \\ [10pt]
		\times \left[ {\left( {p^{0}-eA^{0}\left( {{\rm {\bf x}},t} \right)} 
			\right)^{2}-\left( {{\rm {\bf p}}-e{\rm {\bf A}}\left( {{\rm {\bf x}},t} 
				\right)} \right)^{2}-m^{2}+e{\rm {\bm {\sigma} \bf {H}}}\left( {{\rm {\bf x}},t} 
			\right)-ie{\rm {\bm {\sigma} \bf {E}}}\left( {{\rm {\bf x}},t} \right)} 
		\right] \times \\ [10pt]
		\times A^{\left( + \right)}\left( {{\begin{array}{*{20}c}
					{\varphi_{c} ({\rm {\bf x}},t)} \hfill \\
				\,\,\,\,\,\,\,\,\,0 \hfill \\
		\end{array} }} \right)=0, \\ 
	\end{array}
\end{equation}
\begin{equation}
	\label{eq36}
	\begin{array}{l}
		\left( {H_{CFW}^{\left( - \right)} \left( t \right)} \right)_{odd} \left( 
		{{\begin{array}{*{20}c}
				\,\,\,\,\,\,\,\,\,\,0 \hfill \\
					{\chi_{c} ({\rm {\bf x}},t)} \hfill \\
		\end{array} }} \right)=\beta_{c} m\dfrac{A^{\left( - \right)}}{\left( 
			{-p^{0}+{\rm {\bm {\sigma \pi} }}\left( {{\rm {\bf x}},t} \right)+eA^{0}\left( 
				{{\rm {\bf x}},t} \right)} \right)^{2}}\times \\ [10pt]
		\times \left[ {\left( {p^{0}-eA^{0}\left( {{\rm {\bf x}},t} \right)} 
			\right)^{2}-\left( {{\rm {\bf p}}-e{\rm {\bf A}}\left( {{\rm {\bf x}},t} 
				\right)} \right)^{2}-m^{2}+e{\rm {\bm {\sigma} \bf {H}}}\left( {{\rm {\bf x}},t} 
			\right)+ie{\rm {\bm {\sigma} \bf {E}}}\left( {{\rm {\bf x}},t} \right)} 
		\right] \times \\ [10pt]
		\times A^{\left( - \right)}\left( {{\begin{array}{*{20}c}
					\,\,\,\,\,\,\,\,\,0 \hfill \\
					{\chi_{c} ({\rm {\bf x}},t)} \hfill \\
		\end{array} }} \right)=0. \\ 
	\end{array}
\end{equation}
In (\ref{eq35}) and (\ref{eq36}), ${\rm {\bf H}}=\mbox{rot}{\rm {\bf A}},\,\,\,{\rm {\bf 
		E}}=-\dfrac{\partial {\rm {\bf A}}}{\partial t}-\nabla A^{0}$ are the magnetic 
and electric fields.

As a result, we have obtained in the CFW representation the equations for 
an electron moving in a time-dependent electromagnetic field.
\begin{equation}
	\label{eq37}
		\begin{array}{l}
	\left[ {\left( {p^{0}-eA^{0}\left( {{\rm {\bf x}},t} \right)} 
		\right)^{2}-\left( {{\rm {\bf p}}-e{\rm {\bf A}}\left( {{\rm {\bf x}},t} 
			\right)} \right)^{2}-m^{2}+e{\rm {\bm {\sigma} \bf {H}}}\left( {{\rm {\bf x}},t} 
		\right)-ie{\rm {\bm {\sigma} \bf {E}}}\left( {{\rm {\bf x}},t} \right)} 
	\right] \times \\ [10pt]
	\times A^{\left( + \right)}\varphi_{c} ({\rm {\bf x}},t)=0,
	\end{array}
\end{equation}
\begin{equation}
	\label{eq38}
	\begin{array}{l}
		\left[ {\left( {p^{0}-eA^{0}\left( {{\rm {\bf x}},t} \right)} 
		\right)^{2}-\left( {{\rm {\bf p}}-e{\rm {\bf A}}\left( {{\rm {\bf x}},t} 
			\right)} \right)^{2}-m^{2}+e{\rm {\bm {\sigma} \bf {H}}}\left( {{\rm {\bf x}},t} 
		\right)+ie{\rm {\bm {\sigma} \bf {E}}}\left( {{\rm {\bf x}},t} \right)} 
	\right] \times \\ [10pt]
	\times A^{\left( - \right)}\chi_{c} ({\rm {\bf x}},t)=0.
\end{array}
\end{equation}
If we denote
\begin{equation}
	\label{eq39}
	\varphi_{FG} ({\rm {\bf x}},t)=A^{\left( + \right)}\varphi_{c} ({\rm {\bf 
			x}},t),
\end{equation}
\begin{equation}
	\label{eq40}
	\chi_{FG} ({\rm {\bf x}},t)=A^{\left( - \right)}\chi_{c} ({\rm {\bf 
			x}},t),
\end{equation}
then we obtain the equations in the Feynman-Gell-Mann 
representation \cite{bib4}. In (\ref{eq39}) and (\ref{eq40}), $\varphi_{FG} 
({\rm {\bf x}},t),\,\,\chi_{FG} ({\rm {\bf x}},t)$ are the upper and lower 
spinors of the bispinor $\psi_{FG} ({\rm {\bf x}},t)$
\begin{equation}
	\label{eq41}
	\psi_{FG} ({\rm {\bf x}},t)=\left( {{\begin{array}{*{20}c}
				{\varphi_{FG} ({\rm {\bf x}},t)} \hfill \\
				{\chi_{FG} ({\rm {\bf x}},t)} \hfill \\
	\end{array} }} \right).
\end{equation}
From here, the direct connection between the Foldy-Wouthuysen and 
Feynman-Gell-Mann representationsis evident. 
Previously \cite{bib1}, a similar connection was established 
for for the motion of an electron in static electromagnetic fields.

On the other hand, the connections (\ref{eq39}) and (\ref{eq40}) allow us to solve the problem of ''excess'' solutions of Feynman-Gell-Mann equations \cite{bib4}.

In Eqs. (\ref{eq37}) and (\ref{eq38}), we can perform a similarity transformation and 
obtain equations using spinor wave functions in the Foldy-Wouthuysen 
representation.
\begin{equation}
	\label{eq42}
	\varphi_{c} =\left( {A^{\left( + \right)}} \right)^{-1}\varphi_{FG} ,
	\end{equation}
\begin{equation}
	\label{eq43}
	\left[ {\left( {A^{\left( + \right)}} \right)^{-1}\left( {\left( 
			{p^{0}-eA^{0}} \right)^{2}-\left( {{\rm {\bf p}}-e{\rm {\bf A}}} 
			\right)^{2}-m^{2}+e{\rm {\bm {\sigma} \bf {H}}}-ie{\rm {\bm {\sigma} \bf {E}}}} 
		\right)A^{\left( + \right)}} \right]\varphi_{c} =0.
\end{equation}
\begin{equation}
\label{eq44}
\chi_{c} =\left( {A^{\left( - \right)}} \right)^{-1}\chi_{FG} ,
\end{equation}
\begin{equation}
\label{eq45}
\left[ {\left( {A^{\left( - \right)}} \right)^{-1}\left( {\left( 
		{p^{0}-eA^{0}} \right)^{2}-\left( {{\rm {\bf p}}-e{\rm {\bf A}}} 
		\right)^{2}-m^{2}+e{\rm {\bm {\sigma} \bf {H}}}+ie{\rm {\bm {\sigma} \bf{E}}}} 
	\right)A^{\left( - \right)}} \right]\chi_{c} =0.
\end{equation}
For stationary states, the similarity transformation preserves the energy 
spectrum of a transformed equation. Therefore, the energy spectra of Eqs. (\ref{eq43}) and (\ref{eq37}) are identical. In this case, the spectra of 
Eqs. (\ref{eq45}) and (\ref{eq38}) are also identical. Taking this into account, 
we can analyze simpler Eqs. (\ref{eq37}), (\ref{eq38}) with the wave functions in the 
Feynman-Gell-Mann representation.


\section{Conclusions}

Using Dirac matrices in the chiral representation, we have for the first 
time obtained the equations of motion for electrons in the Foldy-Wouthuysen 
representation in the presence of time-dependent electromagnetic fields. The 
obtained equations can be used in non-perturbative quantum electrodynamics 
with strong electromagnetic fields.

In this work, the connection between the Foldy-Wouthuysen and 
Feynman-Gell-Mann representations has been established in the general case 
of electromagnetic fields. Previously, such a connection was established in 
the case of static electromagnetic fields \cite{bib1}.

In the general case of dynamic electromagnetic fields, the wave function in 
the Foldy-Wouthuysen representation is determined by formulas (\ref{eq19}) and (\ref{eq20})
\begin{equation}
	\label{eq46}
	\psi_{CFW}^{\left( + \right)} ({\rm {\bf x}},t)=\left( 
	{{\begin{array}{*{20}c}
				{\varphi_{c} ({\rm {\bf x}},t)} \hfill \\
				\,\,\,\,\,\,\,0 \hfill \\
	\end{array} }} \right),
\end{equation}
\begin{equation}
	\label{eq47}
	\psi_{CFW}^{\left( - \right)} ({\rm {\bf x}},t)=\left( 
	{{\begin{array}{*{20}c}
				\,\,\,\,\,\,\,\,0 \hfill \\
				{\chi_{c} ({\rm {\bf x}},t)} \hfill \\
	\end{array} }} \right).
\end{equation}
In the case of static electromagnetic fields,
\begin{equation}
	\label{eq48}
	\left( {\psi_{CFW}^{\left( + \right)} ({\rm {\bf x}},t)} \right)_{st} 
	=\left( {{\begin{array}{*{20}c}
				{\varphi_{c}^{st} ({\rm {\bf x}})e^{-i\varepsilon t}} \hfill \\
			\,\,\,\,\,\,\,\,\,\,\,0 \hfill \\
	\end{array} }} \right),\,\,\,\varepsilon >0,
\end{equation}
\begin{equation}
	\label{eq49}
	\left( {\psi_{CFW}^{\left( - \right)} ({\rm {\bf x}},t)} \right)_{st} 
	=\left( {{\begin{array}{*{20}c}
				\,\,\,\,\,\,\,\,\,\,\,0 \hfill \\
				{\chi_{c}^{st} ({\rm {\bf x}})e^{-i\varepsilon t}} \hfill \\
	\end{array} }} \right),\,\,\,\varepsilon <0.
\end{equation}
The set of basic eigenfunctions $\left( {{\begin{array}{*{20}c}
			{\varphi_{c}^{st} ({\rm {\bf x}})e^{-i\varepsilon t}} \hfill \\
			\,\,\,\,\,\,\,\,\,\,\,0 \hfill \\
\end{array} }} \right),\,\,\left( {{\begin{array}{*{20}c}
		\,\,\,\,\,\,\,\,\,\,\,	0 \hfill \\
			{\chi_{c}^{st} ({\rm {\bf x}})e^{-i\varepsilon t}} \hfill \\
\end{array} }} \right)$ and eigenvalues $\varepsilon_{k} $ includes all the 
solutions (\ref{eq48}) and (\ref{eq49}).

The obtained solutions (\ref{eq46}) and (\ref{eq47}) show how to decompose time-dependent wave 
functions into basis functions defined by the expressions (\ref{eq48}) and (\ref{eq49}). The 
solutions (\ref{eq46}), (\ref{eq47}) are unambiguously related to the solutions of 
Feynman-Gell-Mannequations.


\section{Acknowledgments}
This study was conducted within the framework of the scientific program of 
the National Center for Physics and Mathematics, section N5 ''Particle 
physics and cosmology. Stage 2023-2025''.

The author thank A.L.Novoselova for the essential technical assistance in 
the preparation of the paper.



\end{document}